\documentclass[12pt]{article}
\usepackage{pdflscape}

\usepackage{amsmath,amssymb,amsthm,amsxtra,overpic,bbm,bm,epsfig,subfigure}
\usepackage{hyperref}
\hypersetup{
    colorlinks=true,
    linkcolor=black,
    filecolor=blue,
    urlcolor=blue,
    citecolor=blue,
}
\usepackage{mathrsfs}
\usepackage{booktabs}
\usepackage{graphicx}
\usepackage{xcolor}
\usepackage{comment}
\usepackage{epstopdf}
\usepackage{float}
\usepackage{cite}
\usepackage{array}
\usepackage{makecell}
\usepackage{enumitem}
\usepackage{tabularx} 

\usepackage{overpic}
\usepackage{fancyhdr}
\usepackage{float}
\usepackage{rotating}
\usepackage{color}

\usepackage{slashed,stmaryrd}

\def\thefootnote{\fnsymbol{footnote}}

\begin{document}

\vspace{0.2cm}
\begin{center}
{\Large\bf {Precise Values of Running Quark and Lepton Masses \\
in the Standard Model}}
\end{center}
\vspace{0.2cm}

\begin{center}
	{\bf Guo-yuan Huang,~$^{a,~b,~c}$}~\footnote{E-mail: huanggy@ihep.ac.cn}
	\quad
	{\bf Shun Zhou~$^{a,~b}$}~\footnote{E-mail: zhoush@ihep.ac.cn (corresponding author)}
	\\
	\vspace{0.2cm}
	{\small $^a$Institute of High Energy Physics, Chinese Academy of
		Sciences, Beijing 100049, China \\
		$^b$School of Physical Sciences, University of Chinese Academy of Sciences, Beijing 100049, China \\
		$^c$Max-Planck-Institut f\"ur Kernphysik, Postfach
		103980, D-69029 Heidelberg, Germany}
\end{center}

\vspace{1.5cm}

\begin{abstract}
The precise values of the running quark and lepton masses $m^{}_f(\mu)$, which are defined in the modified minimal subtraction scheme ($\overline{\rm MS}$) with $\mu$ being the renormalization scale and the subscript $f$ referring to all the charged fermions in the Standard Model (SM), are very useful for the model building of fermion masses and flavor mixing and for the precision calculations in the SM or its new-physics extensions. In this paper, we calculate the running fermion masses by taking account of the up-to-date experimental results collected by Particle Data Group and the latest theoretical higher-order calculations of relevant renormalization-group equations and matching conditions in the literature. The emphasis is placed on the quantitative estimation of current uncertainties on the running fermion masses, and the linear error propagation method is adopted to quantify the uncertainties, which has been justified by the Monte-Carlo simulations. We identify two main sources of uncertainties, i.e., one from the experimental inputs and the other from the truncations at finite-order loops. The correlations among the uncertainties of running parameters can be remarkable in some cases. The final results of running fermion masses at several representative energy scales are tabulated for further applications.
\end{abstract}

\def\thefootnote{\arabic{footnote}}
\setcounter{footnote}{0}
\newpage

\section{Introduction}

The exciting discovery of the Higgs boson at the CERN Large Hadron Collider (LHC) in 2012 has completed the Standard Model of particle physics (SM) \cite{Aad:2012tfa,Chatrchyan:2012xdj}. However, there remain several fundamentally important issues that cannot be accommodated within the SM framework, i.e., tiny neutrino masses, possible candidates for dark matter and the naturalness problem. No significant deviations from the SM have been found at the energy frontier explored by the LHC, indicating that the energy scale of new physics is probably lying above $\mu \sim 1~{\rm TeV}$.
Although the SM has to be extended to solve those fundamental issues, it can still stand as a successful low-energy effective theory of a complete theory at some high energy scale, e.g., the Grand Unification Theory (GUT) at typically $\Lambda^{}_{\rm GUT} \sim 2 \times 10^{16}~{\rm GeV}$. Any complete theories at high energy scales should be able to reproduce the low-energy observables that are in accordance with the SM predictions.

A convenient and efficient approach was suggested by Steven Weinberg a long time ago to connect the physical parameters in the high-energy full theory to those in the low-energy effective theory~\cite{Weinberg:1980wa,Hall:1980kf}. The basic idea is to integrate out heavy degrees of freedom from the full theory at the decoupling mass scale $\mu^{}_0$ and construct an effective theory at $\mu \ll \mu^{}_0$, where heavy particles just disappear but nonrenormalizable operators should be taken into account. There are several practical advantages of this approach to the multi-scale field theories. First, the renormalized physical parameters in both full and effective theories can be defined without any ambiguities in the modified minimal subtraction ($\overline{\rm MS}$) scheme. Second, the renormalization-group equations (RGEs) of the running parameters can be calculated in a much simpler way than in other mass-dependent renormalization schemes. Third, the connection between the full and effective theories is simply represented as matching conditions at the decoupling scale. Finally, such a procedure is perfectly applicable to a wide class of field theories, whenever two well separated mass scales in question can be identified. In the present work, we apply this approach to the SM and focus on the running fermion masses.

Within the SM, the physical parameters are usually measured at different energy scales. For instance, the electromagnetic fine-structure constant $\alpha$ is experimentally extracted from low-energy processes in which the momentum transfer of photons is vanishingly small, and the corresponding value will be denoted as $\alpha^{}_0$ hereafter. The Fermi constant $G^{}_{\rm F}$ is precisely determined at very low energy transfer from the muon decay data. The strong coupling constant $\alpha^{}_{s}$ has been measured at many different energy scales, but commonly provided at the pole mass of $Z$-boson $M^{}_Z = 91.1876~{\rm GeV}$ via the RGE running as in the review by Particle Data Group (PDG)~\cite{Zyla:2020}. These parameters are always entangled with each other in a complicated manner when running from one energy scale to another. A complete collection of running parameters, especially the running masses, at the concerned energy scale is demanding~\cite{Xing:2019vks}. Over the last two decades, there have been several systematic calculations of the running parameters at various energy scales in the SM~\cite{Fusaoka:1998vc,Xing:2007fb,Xing:2011aa,Antusch:2013jca,Deppisch:2018flu}, which have been proved to be very useful not only for the model building but also for the SM precision physics. In this paper, we make a comprehensive update on the running parameters. The primary motivation for such an update is three-fold. First, tremendous progress has been made in the determination of light quark masses from low-energy data, in particular the sophisticated calculations in the lattice Quantum Chromodynamics (QCD)~\cite{Aoki:2019cca}\footnote{The mass ratios of light quarks are usually extracted from the precise measurements of the $\pi$- and $K$-meson masses within the framework of chiral perturbation theory, whereas the absolute masses of light quarks can be determined either from the Lattice QCD simulations of hadronic mass spectra or from spectral function sum rules for hadronic correlation functions. See, e.g. Ref.~\cite{Zyla:2020}, for a recent review.}. Second, as more and more data have been accumulated, the electroweak observables are measured more and more precisely, e.g., the pole mass of the Higgs boson $M^{}_{h} = 125.10 (14)~{\rm GeV}$ and that of the top quark $M^{}_{t} = 172.4 (7)~{\rm GeV}$ have been found by PDG with an unprecedentedly high precision, where the last digits in parentheses are the standard deviations.
Third, the matching conditions between the full SM with the gauge symmetry ${\rm SU}(3)^{}_{\rm c} \times {\rm SU}(2)^{}_{\rm L} \times {\rm U}(1)^{}_{\rm Y}$ and the effective field theory (EFT) with the unbroken gauge symmetry ${\rm SU}(3)^{}_{\rm c} \times {\rm U}(1)^{}_{\rm EM}$ have been computed up to the two-loop order~\cite{Martin:2018yow}. The matching between the full SM with the low-energy ${\rm SU}(3)^{}_{\rm c} \times {\rm U}(1)^{}_{\rm EM}$ effective theory with massive fermions has been ignored in the previous works~\cite{Fusaoka:1998vc,Xing:2007fb,Xing:2011aa,Antusch:2013jca,Deppisch:2018flu}.

Furthermore, the overall uncertainties on the running parameters, which are also useful for the model building when theoretical predictions are confronted with low-energy observables, should be treated in a consistent way. On the one hand, the RGEs obtained with perturbative computations cannot be exact and will be always terminated at finite orders, leading to additional uncertainties in the evaluation of running parameters.
On the other hand, the uncertainties of different running parameters obtained at a given energy scale could partly originate from a common source of error, e.g., the input value of $\alpha^{}_{s}(\mu)$ at $\mu = M^{}_{Z}$; therefore the correlation can help in reducing the total degree of uncertainties. To evaluate the running parameters at various interesting energy scales, we adopt the recently released code \texttt{SMDR} for the numerical computations~\cite{Martin:2019lqd}.
Other codes for similar purposes, including \texttt{RunDec} \cite{Chetyrkin:2000yt,Herren:2017osy} and \texttt{mr} \cite{Kniehl:2016enc}, are also publicly available.

The rest of this work is structured as follows. In Sec.~2, we describe in detail the general calculational framework, including the inputs at low energies, the running and matching routines, and our numerical method to propagate the uncertainties of various types. Then, the final results of running parameters are presented in Sec.~3. We summarize our main conclusions in Sec.~4.

\section{General Strategy}

\subsection{Theoretical framework}

In the SM with the full gauge symmetry ${\rm SU}(3)^{}_{\rm c} \times {\rm SU}(2)^{}_{\rm L} \times {\rm U}(1)^{}_{\rm Y}$, there are totally 14 independent parameters, which are collectively denoted as follows
\begin{eqnarray} \label{eq:Ya}
\mathcal{Y}^{}_{\rm sm} = \left\{g^{}_{s},~g,~g^{\prime},~v,~\lambda,~y^{}_{t},~y^{}_{b}, ~y^{}_{c}, ~y^{}_{s}, ~y^{}_{d},
~y^{}_{u}, ~y^{}_{\tau}, ~y^{}_{\mu}, ~y^{}_{e}\right\}
\end{eqnarray}
at a given renormalization scale $\mu$ for $\mu > \Lambda^{}_{\rm EW}$ with $\Lambda^{}_{\rm EW}$ being the energy scale of spontaneous gauge symmetry breaking. Some explanations for these parameters are in order.
The SM gauge couplings $g^{}_{s}$, $g$ and $g^{\prime}$ correspond to the ${\rm SU}(3)^{}_{\rm c}$, ${\rm SU}(2)^{}_{\rm L}$ and ${\rm U}(1)^{}_{\rm Y}$ gauge symmetries, respectively. In the Higgs sector, the vacuum expectation value $v$ and the quartic coupling $\lambda$ are chosen to be independent, so the quadratic coupling $m^2$ in the scalar potential can be expressed in terms of $v$ and $\lambda$. In addition, the Yukawa coupling for each fermion $f$ in the SM is denoted as $y^{}_{f}$. After the spontaneous breaking of the electroweak gauge symmetry, i.e., ${\rm SU}(3)^{}_{\rm c} \times {\rm SU}(2)^{}_{\rm L} \times {\rm U}(1)^{}_{\rm Y} \to {\rm SU}(3)^{}_{\rm c} \times {\rm U}(1)^{}_{\rm EM}$, there are 16 physical parameters at low energies that can be found in the global-fit analysis from PDG~\cite{Zyla:2020}:
\begin{eqnarray} \label{eq:mqini}
& & \alpha^{}_{ s}(M^{}_{Z}),
~\alpha^{-1}_{0},
~M^{}_{t},
~m^{}_{b}(m^{}_{b}),
~m^{}_{c}(m^{}_{c}) ,
~m^{}_{s}(2{\rm ~GeV}),
~m^{}_{d}(2{\rm ~GeV}) ,
~m^{}_{u}(2{\rm ~GeV}) ,
~M^{}_{\tau} ,
~M^{}_{\mu} ,
~M^{}_{e} , \notag\\
& & G^{}_{\rm F},
~M^{}_{Z},
~M^{}_{h} ,
~M^{}_{W},
~\sin^2{\theta^{}_{\rm W}} \; .
\end{eqnarray}
Some comments on the experimental measurements of these parameters are helpful. The strong coupling constant $\alpha^{}_{s}(M^{}_{Z})$ is determined by combining the experimental data collected at different energy scales. The fine-structure constant $\alpha^{}_{0}$ is precisely derived from the measurement of $e^{\pm}_{}$ anomalous magnetic moment \cite{Mohr:2015ccw}. The pole masses of heavy SM particles, i.e.,~$M^{}_{t}$, $M^{}_{Z}$, $M^{}_{h}$ and $M^{}_{W}$, are extracted from the data accumulated at high-energy lepton and hadron colliders. It should be noticed that the quoted pole masses $M^{}_{Z}$ and $M^{}_{W}$ from PDG are actually the Breit-Wigner masses, which are defined as $(M^2_{\rm pole} + \Gamma^2)^{1/2}$ with $M^{}_{\rm pole}$ being the true pole mass and $\Gamma$ being the total decay width. The pole mass of $\tau$-lepton is obtained from various lepton collider experiments, while those of light-flavor charged leptons are determined from atomic physics, i.e., the mass ratio of electron to the nucleus in the carbon ions and the mass ratio of muon to electron in the muonic atoms. In addition, the weak mixing angle $\sin^2{\theta^{}_{\rm W}}$ is pinned down from experiments at different energy scales, such as the collider experiments running at the $Z$ pole, the neutrino-nucleon scattering and the atomic parity violation. It is worthwhile to mention that the running of flavor mixing parameters in the quark sector has been ignored in our calculations, as its impact on the running fermion masses is insignificant either in the full SM  above or in the EFTs below the electroweak scale.

The tree-level relations between the fundamental SM parameters in Eq.~(\ref{eq:Ya}) and the low-energy observables in Eq.~(\ref{eq:mqini}) are straightforward. First, the fermion masses and their Yukawa couplings are simply related by $m^{}_{f} = y^{}_{f}v/\sqrt{2}$. Then, the strong coupling constant and the gauge coupling are linked by definition as $\alpha^{}_{s} = g^{2}_{s}/{4\pi}$. The Higgs mass $M^{}_{h} \simeq 125~{\rm GeV}$ can be used to fix the quartic coupling $\lambda \simeq 0.129$ via $M^{}_{h} = \sqrt{2\lambda} v$, given the vacuum expectation value $v \simeq 246~{\rm GeV}$. Finally, the remaining tree-level relations are as follows
\begin{eqnarray} \label{eq:mqini2}
\alpha^{}_{0} = \frac{g^2 g^{\prime 2}}{4\pi \left(g^2 + g^{\prime 2}\right)} \; , \quad G^{}_{\rm F} = \frac{1}{\sqrt{2}v^2} \;, \quad M^{}_{Z} = \frac{v}{2}\sqrt{g^2+g^{\prime 2}} \; , \quad M^{}_{W} = \frac{g v}{2} \; , \quad \sin^{2}\theta^{}_{\rm W} =  \frac{g^{\prime 2}_{}}{g^2+g^{\prime 2}} \; , \quad
\end{eqnarray}
where only three parameters $g$, $g^{\prime}$ and $v$ are involved in the above five observables. According to Table~10.2, Table~10.4 and Fig.~10.2 of Ref.~\cite{Zyla:2020}, the values of  $\sin^2{\theta^{}_{\rm W}}$ and $M^{}_{W}$  derived from $\alpha^{}_{0}$, $G^{}_{\rm F}$ and $M^{}_{Z}$ are much more precise than direct measurements of them. Therefore, we will discard $\sin^2{\theta^{}_{\rm W}}$ and $M^{}_{W}$, and choose $\alpha^{}_{0}$, $G^{}_{\rm F}$ and $M^{}_{Z}$ as basic numerical inputs to fix $g$, $g^{\prime}$ and $v$. The complete set of totally 15 input parameters to be used in our numerical calculations is
\small
\begin{eqnarray} \label{eq:mqini3}
\mathcal{I}^{}_{\rm sm}&  = & \{\alpha^{}_{ s}(M^{}_{Z}) \;,~
\alpha^{}_{0} \;,~
M^{}_{t} \;,~
m^{}_{b}(m^{}_{b}) \;,~
m^{}_{c}(m^{}_{c}) \; ,~
m^{}_{s}(2{\rm ~GeV}) \;, ~
m^{}_{d}(2{\rm ~GeV}) \; ,~
m^{}_{u}(2{\rm ~GeV}) \; ,~
M^{}_{\tau} \; ,~
M^{}_{\mu} \; ,~
 \quad \\ \notag
& &  M^{}_{e} \; , ~
G^{}_{\rm F} \;,~
M^{}_{Z} \;,~
M^{}_{h} \;,~
\Delta \alpha^{(5)}_{\rm had}(M^{}_{Z})
\} \;. \quad
\end{eqnarray}
\normalsize
Following Ref.~\cite{Martin:2019lqd}, we treat $\Delta \alpha^{(5)}_{\rm had}(M^{}_{Z})$, the non-perturbative hadronic radiative contributions to the fine-structure constant $\alpha$, as an input parameter due to its non-perturbative nature. Let us explicitly write down the input parameters with their best-fit values and $1\sigma$ errors ~\cite{Zyla:2020}:
\small
\begin{alignat}{4} \label{eq:inputs}
& \alpha^{}_{ s}(M^{}_{Z}) = 0.1179(10),\,\,\,
&& \alpha^{-1}_{0} = 137.035999084(21),\,\,\,
&&G^{}_{\rm F} = 1.1663787(6)\times 10^{-5}~{\rm GeV^{-2}},\, \notag \\
& M^{}_{Z} = 91.1876(21)~{\rm GeV},\,\,\,
&&M^{}_{h} = 125.10(14)~{\rm GeV} ,\,\,\,
&&M^{}_{t} = 172.4(7)~{\rm GeV},\, \notag \\
&m^{}_{b}(m^{}_{b}) = 4.18(2)~{\rm GeV},\,\,\,
&&m^{}_{c}(m^{}_{c}) = 1.27(2)~{\rm GeV},\,\,\,
&&m^{}_{s}(2{\rm ~GeV}) = 0.093(8)~{\rm GeV},\, \notag \\
&m^{}_{d}(2{\rm ~GeV}) = 4.67(32)~{\rm MeV},\,\,\,
&&m^{}_{u}(2{\rm ~GeV}) = 2.16(38)~{\rm MeV},\,\,\,
&&M^{}_{\tau} = 1.77686(12)~{\rm GeV},\,  \notag \\
& M^{}_{\mu} = 0.1056583745(24)~{\rm GeV},\,\,\,
&&M^{}_{e} = 0.5109989461(31)~{\rm MeV},\,\,\,
&&\Delta \alpha^{(5)}_{\rm had}(M^{}_{Z}) = 0.02764(7)\;.
\end{alignat}
\normalsize
Those quoted errors have been symmetrized and assumed to be Gaussian distributed. We can clearly observe that the uncertainties of inputs are mainly coming from the hadronic sector, i.e., the strong coupling constant $\alpha^{}_{s}$, the quark masses $m^{}_{q}$ and the magnitude of hadronic correction $\Delta \alpha^{(5)}_{\rm had}(M^{}_{Z})$.

\begin{figure*}
	\begin{center}
		\scriptsize
		\begin{overpic}[width=0.95\textwidth]{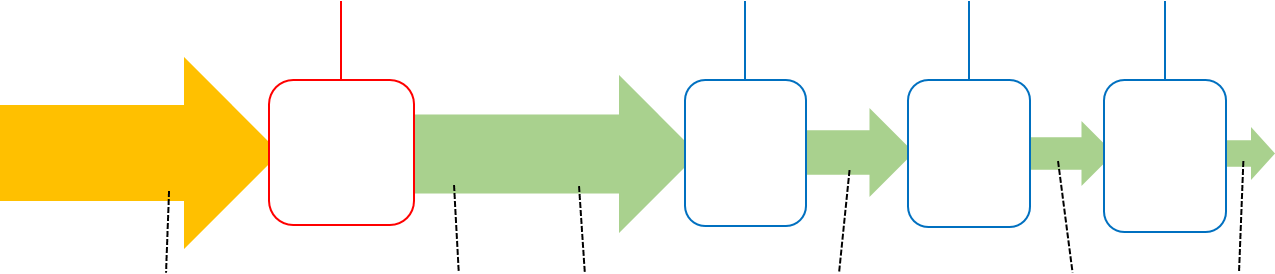}
			\put(1,8.5) {\normalsize$\mathcal{Y}^{}_{\rm sm}$}
			\put(5,9.5) {\textrm{RGEs of full SM}}
			\put(6,7.7) {\cite{Cheng:1973nv,Machacek:1983fi,Machacek:1984zw,Machacek:1983tz,Arason:1991ic,Ford:1992mv,Luo:2002ey,Tarasov,Mihaila:2012fm,Chetyrkin:2012rz,Bednyakov:2012rb,Bednyakov:2012en,
					Chetyrkin:2013wya,Bednyakov:2013eba,Bednyakov:2013cpa,Bednyakov:2014pia,Herren:2017uxn,Davies:2019onf}}
			\put(34,9.5) {\textrm{RGEs of QCD$\times$QED}}
			\put(37,7.8) {\cite{Baikov:2016tgj,Luthe:2017ttc,Herzog:2017ohr,Baikov:2014qja,Luthe:2016xec,Baikov:2017ujl,Kataev:1992dg,Surguladze:1996hx,Erler:1998sy,Mihaila:2014caa,Bednyakov:2016onn}}
			\put(23,10.5) {$t,~h,~Z,~W$}
			\put(22.8,9) {\textrm{decoupling}}
			\put(22,7.1) {\cite{Martin:2018yow,Schroder:2005hy,Chetyrkin:2005ia,Bednyakov:2014fua,Liu:2015fxa}
			}
			\put(55,10.5) {$b$-\textrm{quark}}
			\put(54.7,9) {\textrm{decoupling}}
			\put(53.8,7.2) {\cite{Schroder:2005hy,Chetyrkin:2005ia,Liu:2015fxa,Martin:2018yow}}
			\put(72,10.5) {$\tau$-\textrm{lepton}}
			\put(72,9) {\textrm{decoupling}}
			\put(73,7.2) {\cite{Martin:2018yow}
			}
			\put(87.3,10.5) {$c$-\textrm{quark}}
			\put(87,9) {\textrm{decoupling}}
			\put(86.7,7.2) {\cite{Schroder:2005hy,Chetyrkin:2005ia,Liu:2015fxa,Martin:2018yow}}
			\put(2.3,-1.5) {$M^{}_{t}$ \cite{Jegerlehner:2003py,Marquard:2015qpa,Kataev:2015gvt,Marquard:2016dcn,Martin:2016xsp}, $M^{}_{h}$ \cite{Martin:2014cxa}, $M^{}_{Z}$ \cite{Jegerlehner:2001fb,Jegerlehner:2002em,Martin:2015rea}}
			\put(2.3,-3.5) {$\alpha^{}_{0}$ \cite{Chetyrkin:1996cf,Degrassi:2003rw,Degrassi:2014sxa}, $G^{}_{\rm F}$ \cite{Degrassi:2012ry,Martin:2019lqd}}
			\put(30,-1.5) {{$\alpha^{}_{s}(M^{}_{Z})$}, {\normalsize$\mathcal{Y}^{}_{\rm eff}$}}
			\put(43,-1.5) {$m^{}_{b}(m^{}_{b})$}
			\put(56,-1.5) {$m^{}_{s},m^{}_{d},m^{}_{u},M^{}_{\tau}$ \cite{Gray:1990yh,Bekavac:2007tk,Kataev:2018gle}}
			\put(83,-1.5) {$m^{}_{c}(m^{}_{c})$}
			\put(93,-1.5) {$M^{}_{e},M^{}_{\mu}$}
			\put(93,-3.3) {\cite{Gray:1990yh,Bekavac:2007tk,Kataev:2018gle}
			}
			\put(24,22) {$\underline{\mu=M^{}_{Z}}$}
			\put(54,22) {$\underline{\mu=m^{}_{b}(m^{}_{b})}$}
			\put(73,22) {$\underline{\mu=M^{}_{\tau}}$}
			\put(87,22) {$\underline{\mu=m^{}_{c}(m^{}_{c})}$}
			\put(3,18) {$(n^{}_{q},n^{}_{l})=(6,3)$}
			\put(35,18) {$(n^{}_{q},n^{}_{l})=(5,3)$}
			\put(61,18) {$(n^{}_{q},n^{}_{l})=(4,3)$}
			\put(78,18) {$(n^{}_{q},n^{}_{l})=(4,2)$}
		\end{overpic}
	\end{center}
	\vspace{0.2cm}
	\caption{The flow chart for numerical calculations. First, the running parameters $\mathcal{Y}^{}_{\rm sm}$ in the full SM are specified as the initial conditions above the electroweak scale. The RGEs of the full SM with $n^{}_q=6$ active quarks and $n^{}_l=3$ active leptons are utilized to evolve those parameters to the electroweak scale $\mu = M^{}_{Z}$. After the decoupling of heavy particles $t$, $h$, $Z$ and $W$ at the electroweak scale, an EFT of QCD$\times$QED with $(n^{}_q, n^{}_l) = (5, 3)$ is obtained. Then, we evolve the running parameters based on the resultant RGEs of the EFT until the next threshold is encountered and similar procedures are carried out again. The intermediate decoupling scales have been explicitly shown as the colored frames and the RGEs in between as green arrows, along which the theoretical inputs together with the references are given.}
	\label{fig:1}
\end{figure*}
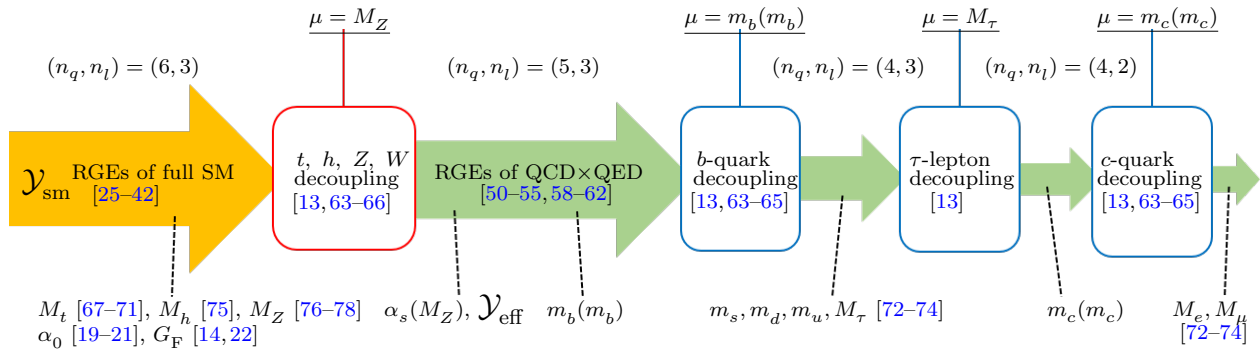

Beyond the tree-level relations, the radiative corrections should be taken into account to link the parameters defined at different renormalization scales. Combining the low-energy inputs $\mathcal{I}^{}_{\rm sm}$ in Eq.~(\ref{eq:inputs}), the RGEs and the matching conditions, one can uniquely determine the SM parameters $\mathcal{Y}^{}_{\rm sm}$ in Eq.~(\ref{eq:Ya}). The theoretical details have been elaborated in the associated paper of the \texttt{SMDR} code \cite{Martin:2019lqd}. Below we briefly summarize the key points.
\begin{itemize}
\item\emph{The fine-structure constant}. The relation between the fine-structure constant at the vanishing momentum transfer $\alpha^{}_0$ and the running gauge couplings defined in the $\overline{\rm MS}$ scheme has been given in Refs.~\cite{Chetyrkin:1996cf,Degrassi:2003rw,Degrassi:2014sxa}, where the contributions up to the three-loop order have been included. To be more specific, the one-loop correction appears in Eq.~(A.2) of Ref.~\cite{Degrassi:2014sxa}, while the higher-order corrections at $\mu = M^{}_{Z}$ can be expressed as the interpolating result, i.e., Eq.~(3.7) in the published version of Ref.~\cite{Degrassi:2014sxa}. Note that $\alpha^{-1}_{0}$ is the most precisely measured parameter of the SM, with a relative precision as high as $10^{-10}$. However, as we will show later, its $\overline{\rm MS}$ value $\alpha(M^{}_{Z})$ at $\mu = M^{}_{Z}$ suffers considerable contamination of radiative corrections, and thus is determined with a relative precision of $\sim 2\times 10^{-4}$.

\item\emph{The Fermi constant}. The Fermi constant measured at low energies can be used to fix the vacuum expectation value $v$ of the Higgs field. Including the loop corrections, it is related to $v$ defined in the Landau gauge via
\begin{eqnarray}
G^{}_{\rm F} = \frac{1+\Delta \widetilde{r}}{\sqrt{2} v^2} \;,
\end{eqnarray}
where $\Delta \widetilde{r} = {\Delta \widetilde{r}^{(1)}}/\left(16\pi^2\right) + {\Delta \widetilde{r}^{(2)}}/\left(16\pi^2\right)^2$ has been written as a sum of one-loop~\cite{Degrassi:2012ry,Martin:2019lqd} and two-loop~\cite{Martin:2019lqd} contributions.
We will use the loop-corrected expectation value $v$ which is defined by the minimum of the Higgs effective potential in the tadpole-free Landau gauge~\cite{Martin:2019lqd},
while the results with the tree-level definition $v^{}_{\rm tree} \equiv \sqrt{-m^2/\lambda}$ can be found in Refs.~\cite{Kniehl:2014yia,Kniehl:2015nwa}.

\item\emph{The RGE running and matching.} In the symmetric phase of the SM, the state-of-the-art RGEs have been implemented in \texttt{SMDR}, including the one-loop, two-loop \cite{Cheng:1973nv,Machacek:1983fi,Machacek:1984zw,Machacek:1983tz,Arason:1991ic,Ford:1992mv,Luo:2002ey} and three-loop~\cite{Tarasov,Mihaila:2012fm,Chetyrkin:2012rz,Bednyakov:2012rb,Bednyakov:2012en,
Chetyrkin:2013wya,Bednyakov:2013eba,Bednyakov:2013cpa,Bednyakov:2014pia,Herren:2017uxn} complete beta functions of gauge couplings, Yukawa couplings and the Higgs quartic coupling and the anomalous dimension of $m^2$,
as well as the independent beta functions for gauge couplings at four-loop order~\cite{Davies:2019onf}.
Other higher-order beta functions are only available with some incomplete leading terms, see e.g. Refs~\cite{Martin:2015eia,Bednyakov:2015ooa,Zoller:2015tha,Chetyrkin:2016ruf}, thus will not be included in our calculation.
Below the electroweak scale, we assume the heavy degrees of freedom in the SM (i.e., the top quark $t$, the Higgs boson $h$, the weak gauge bosons $Z$ and $W^\pm$) to simultaneously decouple from the theory. After the decoupling of these heavy particles, a low-energy EFT with five quarks, three generations of leptons, photon and gluons can be constructed, where the QCD gauge symmetry ${\rm SU}(3)^{}_{\rm c}$ and the gauge symmetry ${\rm U}(1)^{}_{\rm EM}$ for Quantum Electrodynamics (QED) are preserved. To match the parameters in the full theory with those in the EFTs, we should utilize the matching conditions to include the threshold effects from the decoupling of heavy particles. In principle, the matching scale or the decoupling scale can be chosen arbitrarily if radiative corrections at all orders are known. However, to avoid large logarithms arising from the heavy particle masses, it is in practice convenient to set the matching scale to be around the heavy particle masses, $\mu \sim \left\{M^{}_{t}, M^{}_{h}, M^{}_{Z}, M^{}_{W}\right\} \in \left( 80\cdots 173\right)~{\rm GeV}$.
In this work, we fix the matching scale at the pole mass of $Z$, i.e., $M^{}_Z = 91.1876~{\rm GeV}$. The uncertainties associated with the choice of the matching scale can be consistently treated as the theoretical errors. After the electroweak matching procedure, the resultant low-energy EFT of QCD$\times$QED involves the following set of parameters
\begin{eqnarray} \label{eq:Yb}
\mathcal{Y}^{}_{\rm eff} = \left\{\alpha^{}_{s},~\alpha,~m^{}_{b},~m^{}_{c},~m^{}_{s},~ m^{}_{d},~
m^{}_{u},~ m^{}_{\tau},~ m^{}_{\mu}, ~m^{}_{e}\right\} \;.
\end{eqnarray}
The electroweak matching conditions between $\mathcal{Y}^{}_{\rm eff}$ in Eq.~(\ref{eq:Yb}) and $\mathcal{Y}^{}_{\rm sm}$  in Eq.~(\ref{eq:Ya}) have been presented in Ref.~\cite{Martin:2018yow}. The RGEs of the physical parameters for the low-energy EFT of QCD$\times$QED have been extensively studied and updated in the literature. The beta functions of $\alpha^{}_{s}$ have been calculated up to five loops, e.g., in Refs.~\cite{vanRitbergen:1997va,Czakon:2004bu,Poole:2019txl,Baikov:2016tgj,Luthe:2017ttc,Herzog:2017ohr}. These five-loop results have been incorporated into the latest version of the \texttt{RunDec} package~\cite{Herren:2017osy} to calculate the running strong coupling and quark masses below the electroweak scale. The anomalous dimensions of quark masses with pure QCD contributions have been updated in Refs.~\cite{Baikov:2014qja,Luthe:2016xec,Baikov:2017ujl,Chetyrkin:1997dh,Vermaseren:1997fq} up to the five-loop order. In Refs.~\cite{Kataev:1992dg,Surguladze:1996hx,Erler:1998sy,Mihaila:2014caa,Bednyakov:2016onn}, one can find the complete three-loop QCD$\times$QED contributions (including pure QCD, pure QED and their mixture) to the beta functions of $\alpha^{}_{s}$ and $\alpha$ and the anomalous dimensions of fermion masses. At low energies, when a fermion mass threshold (e.g., $b$, $c$ or $\tau$) is crossed, the matching of running parameters between two successive EFTs needs to be performed explicitly. The matching conditions for the gauge couplings and fermion masses are usually given in the form
$
\alpha^{(n^{}_{f}-1)}_{s}(\mu) = \zeta^2_{g^{}_{s}} \alpha^{(n^{}_{f})}_{s}(\mu)
$,
$
\alpha^{(n^{}_{f}-1)}(\mu) = \zeta^2_{e} \alpha^{(n^{}_{f})}(\mu)
$
and
$m^{(n^{}_{f}-1)}_{f} = \zeta^{}_{f} m^{(n^{}_{f})}_{f}
$,
where $\zeta^2_{g^{}_{s}}$, $\zeta^2_{e}$ and $\zeta^{}_{f}$ are the decoupling constants and $n^{}_f$ refers to the number of active fermions in the corresponding EFTs. The expressions of $\zeta^2_{g^{}_{s}}$ and $\zeta^{}_{q}$ (i.e., $\zeta^{}_f$ for quarks) are already known up to the order of four loops for the pure QCD contributions~\cite{Schroder:2005hy,Chetyrkin:2005ia,Liu:2015fxa}. However, the other contributions, e.g. the mixed QCD$\times$QED one, are obtained in Ref.~\cite{Martin:2018yow,Bednyakov:2014fua} only to the two-loop order.

\item\emph{The pole mass and the $\overline{\rm MS}$ running mass.} To derive the $\overline{\rm MS}$ running mass from the pole mass, or vice versa, the relations between the masses in the on-shell scheme and those in the $\overline{\rm MS}$ scheme must be known. In our analysis, the conversions will be carried out from the pole masses of $t$~\cite{Jegerlehner:2003py, Marquard:2015qpa, Kataev:2015gvt, Marquard:2016dcn, Martin:2016xsp}, $\tau$, $\mu$ and $e$ \cite{Gray:1990yh,Bekavac:2007tk,Kataev:2018gle} to their $\overline{\rm MS}$ running masses, while the pole masses of $h$ \cite{Martin:2014cxa} and $Z$ \cite{Jegerlehner:2001fb, Jegerlehner:2002em, Martin:2015rea} have been used to determine the relevant $\overline{\rm MS}$ running couplings.
\end{itemize}
With all the above information, we can compute the $\overline{\rm MS}$ running parameters at a given energy scale, i.e., $\mathcal{Y}^{}_{\rm sm}(\mu)$ and $\mathcal{Y}^{}_{\rm eff}(\mu)$, based on the physical inputs of $\mathcal{I}^{}_{\rm sm}$. A sketch of our computational procedure has been shown in Fig.~\ref{fig:1}. The \texttt{SMDR} code is primarily intended for running the parameters in the full SM at high energies to those in the EFTs at low energies. To conversely obtain the $\overline{\rm MS}$ parameters in the full SM from the low-energy input parameters, a fitting routine has been adopted in \texttt{SMDR}, leading to the determination of $\overline{\rm MS}$ parameters with quite a high precision.

\subsection{Loop truncation and error propagation}

As mentioned before, we can estimate the uncertainties on the running quark and lepton masses with all the precision data available. The error from the experimental inputs is not the only source of uncertainties. Another source, which we shall investigate in the present work, is the theoretical error coming from the perturbation calculations truncated at finite-order loops. More explicitly, the RGEs, the matching conditions and the conversions from on-shell masses to $\overline{\rm MS}$ running masses have been given only at some finite order of perturbations. This type of theoretical error has not been considered in Refs.~\cite{Fusaoka:1998vc,Xing:2007fb,Xing:2011aa,Antusch:2013jca,Deppisch:2018flu}. To quantify this theoretical error, we take the difference between the result obtained by using the highest-order formulas partially from QCD and QED contributions and that by using the formulas completely given at the order lower by one.\footnote{For instance, in the full SM above the electroweak scale, to truncate the three-loop order we compute the running parameters twice, namely, once with the highest three-loop RGEs and once with the two-loop RGEs. The difference between the outcome of two calculations will then be estimated as the uncertainties caused by the finite-loop RGEs~\cite{Bednyakov:2015sca}.}
The loop orders of our truncation procedure for various sources have been summarized in Table~\ref{table:truncationList}.
Some comments and clarifications are helpful.
\begin{itemize}
\item For the RGEs in the EFT of QCD$\times$QED below the electroweak scale, the highest complete RGEs are of three-loop order, including the pure QCD part of $\mathcal{O}(\alpha^{3}_{s})$, the mixed QCD$\times$QED of the order $\mathcal{O}(\alpha^{2}_{s}\alpha) + \mathcal{O}(\alpha^{}_{s}\alpha^{2})$ and the pure QED contribution of the order $\mathcal{O}(\alpha^{3})$. The pure QCD contribution (i.e., without the mixed QCD$\times$QED terms) has actually been calculated up to the five-loop order. Since $\alpha^{}_{s} \gg \alpha$ holds below the electroweak scale, we adopt the five-loop results of the pure QCD contribution and the contributions other than the pure QCD part at the three-loop order.

For the RGEs above the electroweak scale, the perturbation calculations should rely on the gauge couplings $g^{2}_{s}$, $g^2$ and $g^{\prime 2}$, the Yukawa couplings $y^{2}_{f}$ and the quartic Higgs coupling $\lambda$. The complete results are of four-loop order for gauge couplings, and three-loop order for other couplings. Some higher-loop contributions are available but incomplete, so we implement the complete results and take the high-order contributions as the theoretical error.

\item The truncation needs to be performed as well for the decoupling of heavy particles. For $t$, $h$, $W^\pm$ and $Z$ to be decoupled at the electroweak threshold, the complete matching conditions are known up to the two-loop order. The pure QCD contribution to the matching condition for the decoupling of $t$ is known up to four-loop order, whereas the complete matching conditions for the decoupling of $b$, $c$ and $\tau$ in the EFT of QCD$\times$QED are given at the two-loop order. The pure QCD results for the $b$ and $c$ decoupling are also available at the four-loop order. Since the available QCD contribution to the matching conditions is more precise than the complete QCD$\times$QED ones, we will truncate the pure QCD contribution at the four-loop order and the others at the two-loop order, respectively.

\item The conversions of the parameters from the on-shell scheme to those in $\overline{\rm MS}$ scheme, including $M^{}_{\tau}$, $M^{}_{\mu}$, $M^{}_{e}$, $M^{}_{h}$, $M^{}_{t}$, $M^{}_{Z}$, $G^{}_{\rm F}$ and $\alpha^{}_{0}$, are treated with the highest-loop results available. More explicitly, for $M^{}_{h}$, the leading three-loop contribution is included. For $M^{}_{t}$, the pure QCD contribution has been calculated up to the four-loop order, while the other contributions are known at the two-loop order. Therefore, we will truncate the QCD part at the four-loop order and the others at the two-loop order. The conversion between the fine structure constant $\alpha^{}_{0}$ and $\alpha(\mu)$ will be performed at the two-loop order.
\end{itemize}
The errors from the experimental inputs and the theoretical errors from the finite-loop truncation will be treated independently. However, the uncertainties from $b$ and $c$ decoupling in pure QCD are based on the same theory, and likewise for the uncertainties from the $M^{}_{\mu}, M^{}_{e} \leftrightarrow m^{}_{\mu}, m^{}_{e}$ conversions, so they are taken to be $100\%$ correlated.

\begin{table}[!t]
	\small
	\centering
	\caption{Summary of the higher-order contributions implemented in the numerical calculations and the corresponding theoretical errors from the RGEs and the matching conditions truncated at finite-order loops. The listed contributions to the errors are assumed to be independent, except that the uncertainties from $b$ and $c$ decoupling in pure QCD and those from the $M^{}_{\mu}, M^{}_{e} \leftrightarrow m^{}_{\mu}, m^{}_{e}$ conversions are taken to be $100\%$ correlated.}
\vspace{0.5cm}
		\begin{tabular*}{\textwidth}{|l@{\extracolsep{\fill}} c l c|  }
    \hline
	EW matching, complete~\cite{Martin:2018yow} & 2-loop & $M^{}_{\tau} \leftrightarrow m^{}_{\tau}$, complete~\cite{Gray:1990yh,Bekavac:2007tk,Kataev:2018gle} & 3-loop \\ 	
	$b$ and $c$ decoupling, pure QCD~\cite{Schroder:2005hy,Chetyrkin:2005ia,Liu:2015fxa} & 4-loop & 	$M^{}_{\mu},M^{}_{e} \leftrightarrow m^{}_{\mu},m^{}_{e}$, complete~\cite{Gray:1990yh,Bekavac:2007tk,Kataev:2018gle} & 2-loop \\
	$b$ and $c$ decoupling, others~\cite{Martin:2018yow} & 2-loop & 	 $M^{}_{h} \leftrightarrow \lambda$, leading~\cite{Martin:2014cxa} & 3-loop \\
    $\tau$ decoupling, pure QED~\cite{Martin:2018yow}& 2-loop & $M^{}_{t} \leftrightarrow y^{}_{t}$, pure QCD~\cite{Jegerlehner:2003py,Marquard:2015qpa,Kataev:2015gvt,Marquard:2016dcn} & 4-loop \\
	RGEs of QCD$\times$QED, pure QCD~\cite{Baikov:2016tgj,Luthe:2017ttc,Herzog:2017ohr,Baikov:2014qja,Luthe:2016xec,Baikov:2017ujl} & 5-loop &$M^{}_{t} \leftrightarrow y^{}_{t}$, others~\cite{Martin:2016xsp} & 2-loop \\
	RGEs of QCD$\times$QED, others~\cite{Surguladze:1996hx,Erler:1998sy,Mihaila:2014caa,Bednyakov:2016onn} & 3-loop  & 	$M^{}_{Z} \leftrightarrow g,g^{\prime}$, complete~\cite{Martin:2015rea} & 2-loop \\
	RGEs of SM, gauge couplings~\cite{Davies:2019onf} & 4-loop & $G^{}_{\rm F} \leftrightarrow v$, complete~\cite{Martin:2019lqd}  & 2-loop \\
	RGEs of SM, others~\cite{Tarasov,Mihaila:2012fm,Chetyrkin:2012rz,Bednyakov:2012rb,Bednyakov:2012en,
	Chetyrkin:2013wya,Bednyakov:2013eba,Bednyakov:2013cpa,Bednyakov:2014pia,Herren:2017uxn} & 3-loop & $\alpha^{}_{0} \leftrightarrow \alpha$~\cite{Degrassi:2014sxa} & 2-loop \\
	 \hline
	\end{tabular*}
	\label{table:truncationList}
	\end{table}
We adopt the approach of linear error propagation to translate the input errors in Eq.~(\ref{eq:inputs}) into those of the outputs $\mathcal{Y}^{}_{\rm sm}$ and $\mathcal{Y}^{}_{\rm eff}$. The $1\sigma$ experimental errors of the outputs can be obtained by varying the inputs within their $1\sigma$ ranges. By switching on one input error $\delta \mathcal{I}^{}_{k}$ at each time, one can figure out the shift in the output
$\delta \mathcal{Y}^{}_{ik} = ({\partial \mathcal{Y}^{}_{i}}/{\partial \mathcal{I}^{}_{k}}) \cdot \delta \mathcal{I}^{}_{k} $, for the $k$-th input and $i$-th output. In this way, one can find the error matrix
\begin{eqnarray}
S^{}_{ij} \equiv \sum^{}_{k} \left(\frac{\partial \mathcal{Y}^{}_{i}}{\partial \mathcal{I}^{}_{k}} \delta \mathcal{I}^{}_{k}\right)
\left(\frac{\partial \mathcal{Y}^{}_{j}}{\partial \mathcal{I}^{}_{k}} \delta \mathcal{I}^{}_{k} \right)\;,
\end{eqnarray}
where the $1\sigma$ error for each output can be identified as $\sigma^{}_{i} \equiv \sqrt{S^{}_{ii}}$, and the non-diagonal terms with $i\neq j$ quantify the correlations among the output errors of different parameters. The error matrix can be normalized to give the correlation matrix $\rho^{}_{ij}\equiv S^{}_{ij}/\sqrt{S^{}_{ii}S^{}_{jj}}$ such that its elements directly reflect the level of correlation among different parameters. We have demonstrated that the estimation via the linear error propagation turns out to be in perfect agreement with the calculation by the Monte-Carlo approach.

\section{Numerical Results}

Following the strategy outlined above, we update the $\overline{\rm MS}$ running parameters~\cite{Fusaoka:1998vc,Xing:2007fb,Xing:2011aa,Antusch:2013jca,Deppisch:2018flu} in the SM at several representative energy scales, including their best-fit values and the inferred $1\sigma$ uncertainties. Below the electroweak scale, in the EFT with five quarks and three leptons, all the heavy particles including $t$, $h$, $Z$ and $W^\pm$ are integrated out, and we calculate the running parameters at two relevant energy scales: (i) $\mu = M^{}_{W}$, $(n^{}_{q},n^{}_{l})= (5,3)$ without $t$, $h$, $Z$ and $W^\pm$;
(ii) $\mu = M^{}_{Z}$, $(n^{}_{q},n^{}_{l})= (5,3)$ without $t$, $h$, $Z$ and $W^\pm$. Above the electroweak scale, the full gauge symmetry of the SM with the number of active fermions $(n^{}_{q},n^{}_{l})= (6,3)$ is preserved and the typical energy scales are chosen to be (i) $\mu = M^{}_{Z}$;
(ii) $\mu = M^{}_{h}$; (iii) $\mu = M^{}_{t}$; (iv) $\mu = 100~{\rm TeV}$;
(v) $\mu = 10^8~{\rm GeV}$; (vi) $\mu = 10^{12}~{\rm GeV}$. To match the fermion masses below the electroweak scale, we introduce the effective running masses above the electroweak scale as $m^{}_{f} \equiv y^{}_{f} v^{}_{\rm F}/\sqrt{2}$ with $v^{}_{\rm F}  \equiv 246~{\rm GeV} \simeq  2^{-1/4}G^{-1/2}_{\rm F}$, with which one can simply obtain the running Yukawa couplings via $y^{}_{f} = \sqrt{2} m^{}_{f}/v^{}_{\rm F}$ from the running masses $m^{}_f$.

\begin{table}[t!]
	\small
	\centering
	\caption{Running quark masses at some representative energy scales, including $ M^{}_{W} = 80.379~{\rm GeV}$ and $M^{}_{Z}= 91.1876~{\rm GeV}$ in the EFT with the exact ${\rm SU}(3)^{}_{\rm c} \times {\rm U}(1)^{}_{\rm EM}$ gauge symmetry and the number of active fermions $(n^{}_{q},n^{}_{l}) = (5,3)$, as well as $M^{}_{Z}= 91.1876~{\rm GeV}$, $M^{}_{h} = 125.10~{\rm GeV}$, $M^{}_{t} = 173.1~{\rm GeV}$, $\mu = 10^5~{\rm GeV}$, $\mu = 10^8~{\rm GeV}$ and $\mu = 10^{12}~{\rm GeV}$ in the full SM. Above the electroweak scale, the effective running masses have been defined as $m^{}_{f} \equiv y^{}_{f} v^{}_{\rm F}/\sqrt{2}$ with $v^{}_{\rm F}  \equiv 246~{\rm GeV}$. For each parameter, we present its best-fit value $\mathcal{Y}$ and its uncertainty in the form of $\mathcal{Y} \pm \sqrt{\delta^{2}_{\rm exp}+\delta^{2}_{\rm trunc}}$, where the experimental error $\delta^{}_{\rm exp}$ is induced by the input uncertainty while the theoretical error $\delta^{}_{\rm trunc}$ by yet-unknown higher-loop corrections.}
\vspace{0.5cm}
\begin{tabular*}{\textwidth}{c@{\extracolsep{\fill}}c@{\extracolsep{\fill}} c@{\extracolsep{\fill}} c@{\extracolsep{\fill}} c@{\extracolsep{\fill}}c@{\extracolsep{\fill}}c@{\extracolsep{\fill}}c@{\extracolsep{\fill}} }
		\hline
		\hline
		$\mu\,/\,{\rm GeV}$ & ${\rm Theory}$ & $m^{}_{t}\,/\,{\rm GeV}$ & $m^{}_{b} \,/\,{\rm GeV}$ & $m^{}_{c} \,/\,{\rm GeV}$ & $m^{}_{s} \,/\,{\rm MeV}$ & $m^{}_{d} \,/\,{\rm MeV}$ & $m^{}_{u} \,/\,{\rm MeV}$ \\
		\hline
		\addlinespace[0.8mm]
		$M^{}_{W} $ & EFT& $\cdots$ & $2.897 \pm 0.026$ & $0.635 \pm 0.018$ & $54.40 \pm 4.71$ & $2.73 \pm 0.19$ & $1.26 \pm 0.22$    \\
		\addlinespace[0.8mm]
		$M^{}_{Z} $ & EFT & $\cdots$ & $2.866 \pm 0.026$ & $0.628 \pm 0.018$ & $53.80 \pm 4.66$ & $2.70 \pm 0.19$ & $1.24 \pm 0.22$   \\ \addlinespace[0.8mm]
		$M^{}_{Z} $ & Full SM & $168.26 \pm 0.75$ & $2.839 \pm 0.026$ & $0.620 \pm 0.017$ & $53.16 \pm 4.61$ & $2.67 \pm 0.19$ & $1.23 \pm 0.21$   \\ \addlinespace[0.8mm]
		$M^{}_{h} $ & Full SM & $165.05 \pm 0.75$ & $2.768 \pm 0.026$ & $0.607 \pm 0.017$ & $52.00 \pm 4.51$ & $2.61 \pm 0.18$ & $1.20 \pm 0.21$    \\ \addlinespace[0.8mm]
		$M^{}_{t} $ & Full SM & $161.98 \pm 0.75$ & $2.702 \pm 0.025$ & $0.594 \pm 0.017$ & $50.90 \pm 4.41$ & $2.56 \pm 0.18$ & $1.18 \pm 0.20$ \\ \addlinespace[0.8mm]
		$10^5$ & Full SM & $123.77 \pm 0.85$ & $1.908 \pm 0.021$ & $0.435 \pm 0.013$ & $37.47 \pm 3.26$ & $1.88 \pm 0.13$ & $0.86 \pm 0.15$
 \\ \addlinespace[0.8mm]
		$10^8$ & Full SM & $102.49 \pm 0.89$ & $1.502 \pm 0.018$ & $0.350 \pm 0.011$ & $30.34 \pm 2.65$ & $1.52 \pm 0.11$ & $0.69 \pm 0.12$   \\ \addlinespace[0.8mm]
		$10^{12}$ & Full SM & $85.07 \pm 0.89$ & $1.194 \pm 0.015$ & $0.283 \pm 0.009$ & $24.76 \pm 2.17$ & $1.24 \pm 0.09$ & $0.56 \pm 0.10$ \\ \addlinespace[0.8mm]
		\hline
		\hline
	\end{tabular*}
	\label{table:quarkMass}
\end{table}

\begin{table}[h!]
	\small
	\centering
	\caption{Running charged-lepton masses at some representative energy scales, including $ M^{}_{W} = 80.379~{\rm GeV}$ and $M^{}_{Z}= 91.1876~{\rm GeV}$ in the EFT with the exact ${\rm SU}(3)^{}_{\rm c} \times {\rm U}(1)^{}_{\rm EM}$ gauge symmetry and the number of active fermions $(n^{}_{q},n^{}_{l}) = (5,3)$, as well as $M^{}_{Z}= 91.1876~{\rm GeV}$, $M^{}_{h} = 125.10~{\rm GeV}$, $M^{}_{t} = 173.1~{\rm GeV}$, $\mu = 10^5~{\rm GeV}$, $\mu = 10^8~{\rm GeV}$ and $\mu = 10^{12}~{\rm GeV}$ in the full SM with the number of active fermions $(n^{}_{q},n^{}_{l}) = (6,3)$.}
\vspace{0.5cm}
	\begin{tabular*}{\textwidth}{c@{\extracolsep{\fill}}c c c c}
		\hline
		\hline
		$\mu\,/\,{\rm GeV}$ & ${\rm Theory}$ & $m^{}_{\tau} \,/\,{\rm GeV}$ & $m^{}_{\mu} \,/\,{\rm GeV}$ & $m^{}_{e} \,/\,{\rm MeV}$   \\
		\hline
		\addlinespace[0.8mm]
		$M^{}_{W} $ & EFT & $1.74826 \pm 0.00012$ & $0.102925 \pm 0.000018$ & $0.48858 \pm 0.00045$   \\  \addlinespace[0.8mm]
		$M^{}_{Z} $ & EFT& $1.74743 \pm 0.00012$ & $0.102877 \pm 0.000018$ & $0.48835 \pm 0.00045$\\ \addlinespace[0.8mm]
		$M^{}_{Z} $ & Full SM & $1.72856 \pm 0.00028$ & $0.101766 \pm 0.000023$ & $0.48307 \pm 0.00045$  \\ \addlinespace[0.8mm]
		$M^{}_{h} $ & Full SM & $1.73369 \pm 0.00020$ & $0.102065 \pm 0.000020$ & $0.48449 \pm 0.00045$ \\ \addlinespace[0.8mm]
		$M^{}_{t} $ & Full SM& $1.73850 \pm 0.00014$ & $0.102347 \pm 0.000019$ & $0.48583 \pm 0.00045$  \\ \addlinespace[0.8mm]
		$10^5 $ & Full SM & $1.78412 \pm 0.00158$ & $0.105015 \pm 0.000095$ & $0.49850 \pm 0.00064$  \\ \addlinespace[0.8mm]
		$10^8 $ & Full SM & $1.77852 \pm 0.00308$ & $0.104681 \pm 0.000183$ & $0.49691 \pm 0.00098$ \\ \addlinespace[0.8mm]
		$10^{12} $ & Full SM & $1.73194 \pm 0.00466$ & $0.101936 \pm 0.000277$ & $0.48388 \pm 0.00139$ \\ \addlinespace[0.8mm]
		\hline
		\hline
	\end{tabular*}
	\label{table:leptons}
\end{table}

The running masses in the $\overline{\rm MS}$ scheme at different renormalization scales have been summarized in Table~\ref{table:quarkMass} for six quarks and Table~\ref{table:leptons} for three leptons.
The other SM parameters, including the gauge coupling constants and the Higgs-related parameters, have been given in Table~\ref{table:gaugeBelow}, Table~\ref{table:gaugeAbove} and Table~\ref{table:higgs}. The values of the parameters that are not well-defined in the specified theory will be denoted as dots. For each parameter we present its best-fit value $\mathcal{Y}$ and its uncertainty in the form of $\mathcal{Y} \pm \sqrt{\delta^{2}_{\rm exp}+\delta^{2}_{\rm trunc}}$, where the experimental error $\delta^{}_{\rm exp}$ is induced by the input uncertainty while the theoretical error $\delta^{}_{\rm trunc}$ by the loop truncation.
\begin{table}[t!]
	\small
	\centering
	\caption{Running gauge couplings at $\mu = M^{}_{W} = 80.379~{\rm GeV}$ and $\mu = M^{}_{Z}= 91.1876~{\rm GeV}$ in the EFT with the ${\rm SU}(3)^{}_{\rm c} \times {\rm U}(1)^{}_{\rm EM}$ gauge symmetry and the number of active fermions $(n^{}_{q},n^{}_{l}) = (5,3)$.}
\vspace{0.5cm}
	\begin{tabular*}{\textwidth}{c@{\extracolsep{\fill}} c c c  }
		\hline
		\hline
		$\mu$ & ${\rm Theory}$ & $\alpha^{}_{s}$ & $\alpha^{-1}$  \\
		\hline
		\addlinespace[0.8mm]
		$M^{}_{W} $ & EFT & $0.1199 \pm 0.0010$ & $127.937 \pm 0.026$   \\ \addlinespace[0.8mm]
		$M^{}_{Z} $ & EFT & $0.1176 \pm 0.0010$ & $127.754 \pm 0.026$  \\ \addlinespace[0.8mm]
		\hline
		\hline
	\end{tabular*}
	\label{table:gaugeBelow}
\end{table}

\begin{table}[t!]
	\small
	\centering
	\caption{Running gauge couplings at some representative energy scales, including $M^{}_{Z}= 91.1876~{\rm GeV}$, $M^{}_{h} = 125.10~{\rm GeV}$, $M^{}_{t} = 173.1~{\rm GeV}$, $\mu = 10^5~{\rm GeV}$, $\mu = 10^8~{\rm GeV}$ and $\mu = 10^{12}~{\rm GeV}$ in the full SM with the number of active fermions $(n^{}_{q},n^{}_{l}) = (6,3)$.}
\vspace{0.5cm}
	\begin{tabular*}{\textwidth}{c@{\extracolsep{\fill}} c c  c c }
		\hline
		\hline
		$\mu$ & ${\rm Theory}$ &  $g^{}_{s}$ & $g$ & $g^{\prime}$  \\
		\hline
		\addlinespace[0.8mm]
		$M^{}_{Z} $ & Full SM & $1.2104 \pm 0.0051$ & $0.65100 \pm 0.00028$ & $0.357254 \pm 0.000069$  \\ \addlinespace[0.8mm]
		$M^{}_{h} $ & Full SM & $1.1855 \pm 0.0048$ & $0.64934 \pm 0.00028$ & $0.357893 \pm 0.000070$  \\ \addlinespace[0.8mm]
		$M^{}_{t} $ & Full SM & $1.1618 \pm 0.0045$ & $0.64765 \pm 0.00028$ & $0.358545 \pm 0.000070$  \\ \addlinespace[0.8mm]
		$10^5~{\rm GeV}$ & Full SM & $0.8711 \pm 0.0019$ & $0.61644 \pm 0.00025$ & $0.372179 \pm 0.000079$  \\ \addlinespace[0.8mm]
		$10^8~{\rm GeV}$ & Full SM & $0.7182 \pm 0.0010$ & $0.58690 \pm 0.00022$ & $0.388806 \pm 0.000091$  \\ \addlinespace[0.8mm]
		$10^{12}~{\rm GeV}$ & Full SM & $0.6017 \pm 0.0006$ & $0.55325 \pm 0.00019$ & $0.414821 \pm 0.000111$ \\ \addlinespace[0.8mm]
		\hline
		\hline
	\end{tabular*}
	\label{table:gaugeAbove}
\end{table}

\begin{table}[!t]
	\small
	\centering
	\caption{Running Higgs parameters at some representative energy scales, including $M^{}_{Z}= 91.1876~{\rm GeV}$, $M^{}_{h} = 125.10~{\rm GeV}$, $M^{}_{t} = 173.1~{\rm GeV}$, $\mu = 10^5~{\rm GeV}$, $\mu = 10^8~{\rm GeV}$ and $\mu = 10^{12}~{\rm GeV}$ in the full SM with the number of active fermions $(n^{}_{q},n^{}_{l}) = (6,3)$.}
\vspace{0.5cm}
	\begin{tabular*}{\textwidth}{c@{\extracolsep{\fill}} c c c c}
		\hline
		\hline
		$\mu$ & ${\rm Theory}$ & $v \,/\,{\rm GeV}$ & $\lambda$ & $-m^2 \,/\,{\rm GeV}^2$   \\
		\hline
		\addlinespace[0.8mm]
		$M^{}_{Z} $ & Full SM & $248.404 \pm 0.036$ & $0.13947 \pm 0.00045$ & $8434 \pm 21$  \\ \addlinespace[0.8mm]
		$M^{}_{h} $ & Full SM & $247.482 \pm 0.023$ & $0.13259 \pm 0.00035$ & $8525 \pm 22$
		\\ \addlinespace[0.8mm]
		$M^{}_{t} $ & Full SM & $246.605 \pm 0.011$ & $0.12607 \pm 0.00030$ & $8612 \pm 23$
		 \\ \addlinespace[0.8mm]
		$10^5~{\rm GeV} $ & Full SM & $236.500 \pm 0.209$ & $0.04993 \pm 0.00239$ & $9740 \pm 35$
		\\ \addlinespace[0.8mm]
		$10^8~{\rm GeV} $ & Full SM  & $232.831 \pm 0.402$ & $0.01525 \pm 0.00408$ & $10213 \pm 38$
		\\ \addlinespace[0.8mm]
		$10^{12}~{\rm GeV} $ & Full SM & $232.510 \pm 0.624$ & $-0.00402 \pm 0.00522$ & $10274 \pm 32$
		\\ \addlinespace[0.8mm]
		\hline
		\hline
	\end{tabular*}
	\label{table:higgs}
\end{table}

As briefly mentioned before, the $\overline{\rm MS}$ running parameters may share the common sources of input uncertainties, so their overall uncertainties are actually correlated.
For further reference and completeness, we summarize in Table~\ref{table:corr2} and Table~\ref{table:corr3} the correlation matrix of those $\overline{\rm MS}$ parameters at $\mu = M^{}_{Z}$ in the EFT of QCD$\times$QED and the full SM, respectively. To make use of these numerical results in confronting the model predictions with observations, one may first reconstruct the error matrix $S^{}_{ij}$ based on the normalized correlation matrix $\rho^{}_{ij}$ via $ S^{}_{ij} = \rho^{}_{ij} \sigma^{}_{i} \sigma^{}_{j}$, and then quantify the statistical significance of deviations by $\chi^2 = (\mathcal{Y}^{}_{i} - \mathcal{Y}^{\rm bf}_{i} )
\left(S^{-1}\right)^{}_{ij}(\mathcal{Y}^{}_{j} - \mathcal{Y}^{\rm bf}_{j} )$ with $\mathcal{Y}^{}_{j}$ being the model prediction and $\mathcal{Y}^{\rm bf}_{j}$ being the given central value.

\begin{figure}[t!]
	\begin{center}
		\hspace{0.0cm}
		\includegraphics[width=0.95\textwidth]{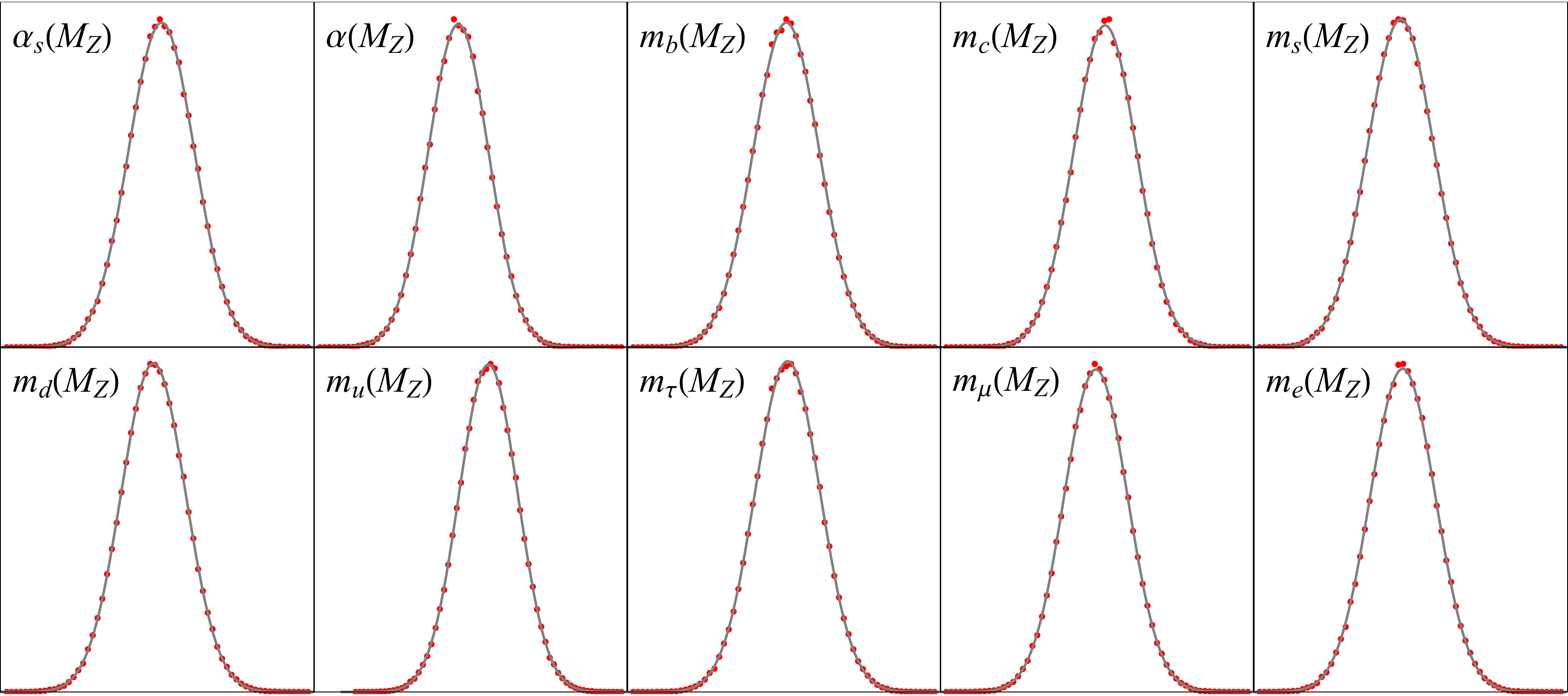}
	\end{center}
	\vspace{-0.4cm}
	\caption{The comparison between the uncertainties of running parameters evaluated by the linear error propagation method (gray curves) and those by the Monte-Carlo simulations (red dots) with $2\times 10^{5}$ sampling points, where the running parameters are given at $\mu = M^{}_{Z}$ in the EFT with the exact ${\rm SU}(3)^{}_{\rm c} \times {\rm U}(1)^{}_{\rm EM}$ gauge symmetry and the number of active fermions $(n^{}_{q},n^{}_{l}) = (5,3)$. For clarity, the labels for the horizontal and vertical axes are not explicitly shown. The gray curves are generated by using the Gaussian distribution with the central values and $1\sigma$ experimental uncertainties, while the red dots stand for the posteriors yielded by the Monte-Carlo simulations.}
	\label{fig:2}
\end{figure}

Now we demonstrate that it is reasonable to implement the linear error propagation by the Monte-Carlo simulations. With $2\times 10^{5}$ sampling points, the posterior distributions of the running parameters have been generated in the EFT of QCD$\times$QED at $\mu = M^{}_{Z}$, and they are compared in Fig.~\ref{fig:2} with the distributions obtained from the linear error propagation.
As one can observe from Fig.~\ref{fig:2}, a perfect agreement between these two methods is found.
Since the uncertainties on the input parameters are at the perturbative level, if their values are updated in the future with new best-fit values and even smaller errors, it is straightforward to recalculate the best-fit values and errors of running parameters by utilizing the error dependence matrix provided in Table~\ref{table:coe2}. For instance, if the best-fit values of input parameters are changed by the amount of $\Delta \mathcal{I}^{}_{k}$, the best-fit values of the running parameters will be accordingly shifted as
$\mathcal{Y}^{\rm new-bf}_{i}  = \mathcal{Y}^{\rm old-bf}_{i} +\sum^{}_{k} ({\partial \mathcal{Y}^{}_{i}}/{\partial \mathcal{I}^{}_{k}}) \cdot \Delta \mathcal{I}^{}_{k} $. Similarly, if the uncertainties on the input parameters are improved, one can determine the $1\sigma$ experimental errors of the running parameters via $\delta \mathcal{Y}^{}_{ik} = ({\partial \mathcal{Y}^{}_{i}}/{\partial \mathcal{I}^{}_{k}}) \cdot \delta \mathcal{I}^{}_{k} $. Such a treatment is valid as long as the uncertainties are small and the linear error propagation is justified, which should be the case for the future measurements with more data.

Apart from the above comments, some important observations from the numerical calculations can be made.
\begin{itemize}
\item In Table~\ref{table:errorSourcesAbove}, we present the fractions of the output uncertainty contributed from each input parameter, which are characterized by $\left({\delta\mathcal{Y}^{}_{ik}}/{\sigma^{}_{i}} \right)^2$, at $\mu = M^{}_{Z}$ in the full SM. For each output parameter the total fraction by definition amounts to $100\%$. For the running masses of five quarks $b$, $c$, $s$, $d$ and $u$ at high energy scales, we identify two major sources of input errors: (i) the strong coupling constant $\alpha^{}_{s}(M^{}_{Z})$; (ii) their input $\overline{\rm MS}$ masses at low energies. For the masses of three light quarks, i.e., $m^{}_{s}$, $m^{}_{d}$ and $m^{}_{u}$, the uncertainties from their input $\overline{\rm MS}$ masses are dominant. A better knowledge of $m^{}_{q}$ and $\alpha^{}_{s}(M^{}_{Z})$ at low energies in the future will thus greatly improve the accuracy of  their running values at high energy scales. For the top quark $t$, its running mass above the electroweak scale is mainly affected by the input of its pole mass.

\item The vacuum expectation value $v$ of the SM Higgs field does not change much at different renormalization scales. As is well known, however, the quartic coupling $\lambda$ has the risk to be negative at the energy scales above $\mu = 10^{8}~{\rm GeV}$, e.g., $\lambda<0$ is within the $4\sigma$ range at $\mu = 10^{8}~{\rm GeV}$ and $\lambda<0$ is the best fit for $\mu = 10^{12}~{\rm GeV}$. New physics beyond the SM has to come into play to save the vacuum from being unstable. For the latest discussions in this aspect, one should be referred to Refs.~\cite{Xing:2011aa,Degrassi:2012ry,Holthausen:2011aa,EliasMiro:2011aa,Bezrukov:2012sa,Buttazzo:2013uya,DiLuzio:2014bua,Bednyakov:2015sca,Andreassen:2014gha}.

\item It is worthwhile to notice that the theoretical uncertainties arising from the truncation of finite loops could be dominant for the electroweak parameters $m^{}_{\mu}$, $m^{}_{e}$, $g$ and $g^{\prime}$. For the other parameters, the theoretical errors are relatively small compared to the experimental ones. The electroweak matching procedure dominates the theoretical errors of the running parameters $g^{}_{s}$, $m^{}_{b}$, $m^{}_{s}$, $m^{}_{d}$ and $m^{}_{u}$ above the electroweak scale.
\end{itemize}

\section{Concluding Remarks}

Although the SM has been experimentally proved to be extremely successful, it must be extended to explain tiny neutrino masses, and to provide the dark matter candidate and the solution to the hierarchy problem. The energy scale for new physics has been pushed by the experiments at the LHC to be beyond TeV. In the present paper, we update the running fermion masses and other $\overline{\rm MS}$ parameters at several representative energy scales, e.g., $\mu = 10^5~{\rm GeV}$, $10^8~{\rm GeV}$ and $10^{12}~{\rm GeV}$, where new physics may come into play. In addition, the running parameters are also evaluated at $\mu = M^{}_Z$, $M^{}_h$ and $M^{}_t$, which will be useful for the precision calculations in the SM.

Compared with the previous similar works in Refs.~\cite{Fusaoka:1998vc,Xing:2007fb,Xing:2011aa,Antusch:2013jca,Deppisch:2018flu}, a number of significant improvements should be emphasized. First, tremendous progress has been made in the theoretical high-order calculations of the RGEs and the matching conditions in the SM and its related EFTs in the past few years. In particular, stimulated by the exciting discovery of the Higgs boson in 2012, the theoretical treatment of radiative corrections from the electroweak sector has been significantly advanced. Using the publicly available code \texttt{SMDR} with the state-of-the-art theoretical knowledge, we have incorporated all the latest results of relevant RGEs and matching conditions. Second, the experimental information has also been greatly changed, as indicated in the latest version of Review of Particle Physics from PDG, which is the source of all the input parameters used in our numerical calculations. Third, we deal with the uncertainties of running parameters in a consistent manner. Both the experimental input errors and the theoretical uncertainties due to the finite-loop calculations  are relevant error sources.

Our main results of the running quark and lepton masses, gauge couplings and scalar parameters are summarized in a number of tables, namely, from Table~\ref{table:quarkMass} to Table~\ref{table:higgs}, which can be directly used for further applications. The linear error propagation has been verified to be an efficient and convenient method in deriving the uncertainties on the running parameters. We find that the theoretical uncertainties due to loop truncations can be dominant for the running parameters $m^{}_{e}$, $m^{}_{\mu}$, $g$ and $g^{\prime}$, whereas for other running parameters the experimental input uncertainties are the major error sources. The results presented here could be easily improved with future update on the experimental and theoretical knowledge.

\section*{Acknowledgments}

The authors are indebted to Prof.~Zhi-zhong Xing for helpful discussions and valuable suggestions. This work was supported in part by the National Natural Science Foundation of China under Grants No.~11775232 and No.~11835013, by the CAS Center for Excellence in Particle Physics, and by the Alexander von Humboldt Foundation.

\bibliographystyle{utcaps_mod}
\bibliography{references}

\newpage
\begin{landscape}
	
	\begin{table}[t]
		\scriptsize
		\centering
		\begin{tabular*}{\linewidth}{c |@{\extracolsep{\fill}} ccccccc c c c}
			\hline
			\hline
			$\rho^{}_{ij}$ at $M^{}_{Z}$& $\alpha^{}_{s}$ & $\alpha^{-1}$ &$m^{}_{b}$ & $m^{}_{c}$ & $m^{}_{s}$ & $m^{}_{d}$ & $m^{}_{u}$ & $m^{}_{\tau}$ & $m^{}_{\mu}$ & $m^{}_{e}$\\
			\hline
			$\alpha^{}_{s}$& 1 & 0.0054 & -0.62 & -0.6 & -0.12 & -0.15 & -0.059 & 0.0029 & 0.0031 & 0.0014 \\
			$\alpha^{-1}$& 0.0054 & 1 & -0.0033 & -0.0032 & -0.00064 & -0.00078 & -0.00031 & 0.048 & 0.03 & 0.0095 \\
			$m^{}_{b}$& -0.62 & -0.0033 & 1 & 0.37 & 0.073 & 0.09 & 0.036 & -0.002 & -0.0022 & -0.00094 \\
			$m^{}_{c}$& -0.6 & -0.0032 & 0.37 & 1 & 0.073 & 0.09 & 0.036 & -0.0017 & -0.0043 & -0.0019 \\
			$m^{}_{s}$& -0.12 & -0.00064 & 0.073 & 0.073 & 1 & 0.018 & 0.0071 & -0.00034 & -0.00038 & -0.00016 \\
			$m^{}_{d}$& -0.15 & -0.00078 & 0.09 & 0.09 & 0.018 & 1 & 0.0088 & -0.00042 & -0.00046 & -0.0002 \\
			$m^{}_{u}$& -0.059 & -0.00031 & 0.036 & 0.036 & 0.0071 & 0.0088 & 1 & -0.00016 & -0.00019 & -8.3e-05 \\
			$m^{}_{\tau}$& 0.0029 & 0.048 & -0.002 & -0.0017 & -0.00034 & -0.00042 & -0.00016 & 1 & 0.0015 & 0.00036 \\
			$m^{}_{\mu}$& 0.0031 & 0.03 & -0.0022 & -0.0043 & -0.00038 & -0.00046 & -0.00019 & 0.0015 & 1 & 1 \\
			$m^{}_{e}$& 0.0014 & 0.0095 & -0.00094 & -0.0019 & -0.00016 & -0.0002 & -8.3e-05 & 0.00036 & 1 & 1 \\
			\hline
			\hline
		\end{tabular*}
		\caption{The error correlation matrix at $\mu = M^{}_{Z}$ in the EFT of ${\rm SU}(3)^{}_{\rm c} \times {\rm U}(1)^{}_{\rm EM}$ with the number of active fermions $(n^{}_{q},n^{}_{l}) = (5,3)$.}
		\label{table:corr2}
	\end{table}
	
	\begin{table}[t]
		\scriptsize
		\centering
		\begin{tabular*}{\linewidth}{c|@{\extracolsep{\fill}}cccccccccccccccc}
			\hline
			\hline
			$\rho^{}_{ij}$ at $M^{}_{Z}$ & $g^{}_{s}$ & $g^{}_{}$ & $g^{\prime}_{}$ & $m^{}_{t}(y^{}_{t})$ & $m^{}_{b}(y^{}_{b})$ & $m^{}_{c}(y^{}_{c})$ & $m^{}_{s}(y^{}_{s})$ & $m^{}_{d}(y^{}_{d})$ & $m^{}_{u}(y^{}_{u})$ & $m^{}_{\tau}(y^{}_{\tau})$ & $m^{}_{\mu}(y^{}_{\mu})$ & $m^{}_{e}(y^{}_{e})$ & $v$ & $\lambda$ & $-m^2$  \\
			\hline
$g^{}_{s}$& 1 & -0.017 & 0.013 & -0.056 & -0.6 & -0.59 & -0.12 & -0.14 & -0.058 & 0.067 & 0.049 & 0.013 & -0.07 & -0.082 & -0.041 \\
$g$& -0.017 & 1 & -0.74 & 0.098 & 0.0093 & 0.0095 & 0.0018 & 0.0022 & 0.00091 & -0.058 & -0.041 & -0.0095 & 0.057 & 0.076 & 0.054 \\
$g^{\prime}_{}$& 0.013 & -0.74 & 1 & -0.095 & -0.0064 & -0.0067 & -0.0013 & -0.0016 & -0.00063 & 0.05 & 0.029 & 0.0043 & -0.065 & -0.074 & -0.053 \\
$m^{}_{t} (y^{}_{t})$& -0.056 & 0.098 & -0.095 & 1 & 0.019 & 0.024 & 0.004 & 0.005 & 0.002 & -0.85 & -0.61 & -0.15 & 0.94 & 0.78 & 0.58 \\
$m^{}_{b}(y^{}_{b})$& -0.6 & 0.0093 & -0.0064 & 0.019 & 1 & 0.37 & 0.074 & 0.091 & 0.037 & -0.023 & -0.018 & -0.005 & 0.027 & 0.039 & 0.017 \\
$m^{}_{c}(y^{}_{c})$& -0.59 & 0.0095 & -0.0067 & 0.024 & 0.37 & 1 & 0.073 & 0.09 & 0.036 & -0.029 & -0.024 & -0.0069 & 0.032 & 0.042 & 0.019 \\
$m^{}_{s}(y^{}_{s})$& -0.12 & 0.0018 & -0.0013 & 0.004 & 0.074 & 0.073 & 1 & 0.018 & 0.0072 & -0.0051 & -0.0038 & -0.001 & 0.0056 & 0.0077 & 0.0034 \\
$m^{}_{d}(y^{}_{d})$& -0.14 & 0.0022 & -0.0016 & 0.005 & 0.091 & 0.09 & 0.018 & 1 & 0.0088 & -0.0063 & -0.0047 & -0.0013 & 0.0069 & 0.0095 & 0.0042 \\
$m^{}_{u}(y^{}_{u})$& -0.058 & 0.00091 & -0.00063 & 0.002 & 0.037 & 0.036 & 0.0072 & 0.0088 & 1 & -0.0025 & -0.0019 & -0.00051 & 0.0028 & 0.0038 & 0.0017 \\
$m^{}_{\tau}(y^{}_{\tau})$& 0.067 & -0.058 & 0.05 & -0.85 & -0.023 & -0.029 & -0.0051 & -0.0063 & -0.0025 & 1 & 0.59 & 0.14 & -0.91 & -0.64 & -0.47 \\
$m^{}_{\mu}(y^{}_{\mu})$& 0.049 & -0.041 & 0.029 & -0.61 & -0.018 & -0.024 & -0.0038 & -0.0047 & -0.0019 & 0.59 & 1 & 0.85 & -0.65 & -0.45 & -0.33 \\
$m^{}_{e}(y^{}_{e})$& 0.013 & -0.0095 & 0.0043 & -0.15 & -0.005 & -0.0069 & -0.001 & -0.0013 & -0.00051 & 0.14 & 0.85 & 1 & -0.16 & -0.11 & -0.081 \\
$v$& -0.07 & 0.057 & -0.065 & 0.94 & 0.027 & 0.032 & 0.0056 & 0.0069 & 0.0028 & -0.91 & -0.65 & -0.16 & 1 & 0.7 & 0.52 \\
$\lambda$& -0.082 & 0.076 & -0.074 & 0.78 & 0.039 & 0.042 & 0.0077 & 0.0095 & 0.0038 & -0.64 & -0.45 & -0.11 & 0.7 & 1 & 0.96 \\
$-m^2$& -0.041 & 0.054 & -0.053 & 0.58 & 0.017 & 0.019 & 0.0034 & 0.0042 & 0.0017 & -0.47 & -0.33 & -0.081 & 0.52 & 0.96 & 1 \\
			\hline
			\hline
		\end{tabular*}
		\caption{The error correlation matrix at $\mu = M^{}_{Z}$ in the full SM with the ${\rm SU}(3)^{}_{\rm c} \times {\rm SU}(2)^{}_{\rm L} \times {\rm U}(1)^{}_{\rm Y}$ gauge symmetry and the number of active fermions $(n^{}_q, n^{}_l) = (6, 3)$. Notice that $m^{}_f(y^{}_f)$ in the first row and column actually refers to the running fermion mass $m^{}_f$ and the corresponding Yukawa coupling $y^{}_f$.}
		\label{table:corr3}
	\end{table}

\end{landscape}

\begin{landscape}

\begin{table}[t]
	\scriptsize
	\centering
		\begin{tabular*}{\linewidth}{c |@{\extracolsep{\fill}} ccccccccccccccccc}
		\hline
		\hline
$\left(\frac{\mathcal{I}^{}_{k}}{\mathcal{Y}^{}_{i}}\right) \cdot \left(\frac{\partial{\mathcal{Y}^{}_{i}}}{\partial{\mathcal{I}^{}_{k}}} \right)$ at $M^{}_{Z}$ & $g^{}_{s}$ & $g^{}_{}$ & $g^{\prime}_{}$ & $m^{}_{t} (y^{}_{t})$ & $m^{}_{b}(y^{}_{b})$ & $m^{}_{c}(y^{}_{c})$ & $m^{}_{s}(y^{}_{s})$ & $m^{}_{d}(y^{}_{d})$ & $m^{}_{u}(y^{}_{u})$ & $m^{}_{\tau}(y^{}_{\tau})$ & $m^{}_{\mu}(y^{}_{\mu})$ & $m^{}_{e}(y^{}_{e})$ & $v$ & $\lambda$ & $-m^2$  \\
\hline
$\alpha^{}_{s}(M^{}_{Z})$& 0.49& -0.00085& 0.00027& -0.025& -0.65& -2& -1.2& -1.2& -1.2& 0.0011& 0.0011& 0.0012& -0.0011& -0.028& -0.011\\
$\alpha^{-1}_{0}$& -2.1e-05& 0.23& -0.74& 0.0015& 0.00068& 0.0063& 0.00073& 0.00073& 0.0056& 0.016& 0.026& 0.045& -3.9e-05& -0.0015& 0.011\\
$m^{}_{b}(m^{}_{b})$& 2.8e-07& -8.6e-06& 2.6e-06& -4.4e-05& 1.2& -0.026& -0.013& -0.013& -0.013& -7e-06& -1.2e-05& -2.2e-05& 3.1e-06& -2e-05& -1.3e-06\\
$m^{}_{c}(m^{}_{c})$& 0& -6.9e-08& 2.1e-08& -4.1e-07& -4.1e-07& 1.4& -4.1e-07& -4.1e-07& -4.1e-07& -4.1e-07& -3.3e-05& -8.1e-05& 4.1e-07& -1.9e-07& 6.7e-07\\
$m^{}_{s}(2~{\rm GeV})$&0& 0& 0& 0& 0& 0& 1& 0 & 0 & 0& 0& 0& 0& 0& 0\\
$m^{}_{d}(2~{\rm GeV})$&0 &  0 &  0 &  0 &  0 &  0 &  0 &1& 0 & 0 & 0& 0& 0& 0& 0\\
$m^{}_{u}(2~{\rm GeV})$& 0& 0&0&0&0&0&0&0 & 1&0& 0 & 0 & 0& 0& 0\\
$M^{}_{\tau}$& 0 & 6e-06& -1.8e-06& 4.6e-06& 4.6e-06& 2.6e-06& 4.6e-06& 4.6e-06& 4.6e-06& 1& -1.7e-05& -4.6e-05& -4.6e-06& 1.2e-05& 2.2e-06\\
$M^{}_{\mu}$& 2.6e-08& 2.3e-07& -1.4e-07& -3.1e-06& 0& -1.8e-08& 1.1e-07& 1.1e-07& 1.1e-07& -8.4e-08& 1& -6.1e-06& 8.2e-08& -1.9e-06& 1.4e-05\\
$M^{}_{e}$& -3e-07& 1.1e-06& -4e-07& 4.5e-05& -1.1e-06& -1.9e-06& -1.3e-06& -1.3e-06& -1.3e-06& -1.8e-06& -1.8e-06& 1& 1.8e-06& 2.4e-05& -0.00015\\
$\Delta \alpha^{(5)}_{\rm had} (M^{}_{Z})$& 3.8e-07& -0.0064& 0.021& -1.4e-05& -2e-05& -0.00018& -2.2e-05& -2.2e-05& -0.00016& -0.00047& -0.00075& -0.0013& 1.5e-06& 6e-05& 7e-05\\
$G^{}_{\rm F}$& 5.2e-05& 0.72& -0.22& 0.49& 0.49& 0.49& 0.49& 0.49& 0.49& 0.49& 0.49& 0.49& -0.49& 1.1& 0.07\\
$M^{}_{Z}$& -5.2e-05& 1.4& -0.42& -0.0022& -0.004& -0.0052& -0.0051& -0.0051& -0.0052& -0.0046& -0.0046& -0.0046& -0.00011& -0.029& -0.024\\
$M^{}_{h}$& 4.3e-05& -0.0011& 0.00032& -0.0028& 0.0022& 0.0022& 0.0022& 0.0022& 0.0022& 0.0022& 0.0022& 0.0022& -0.0022& 1.7& 1.8\\
$M^{}_{t}$& -0.0073& 0.01& -0.0045& 1& -0.02& -0.033& -0.033& -0.033& -0.033& -0.032& -0.032& -0.032& 0.032& 0.59& 0.34\\
		\hline
		\hline
	\end{tabular*}
	\caption{The error dependence coefficients at $\mu = M^{}_{Z}$ in the full SM, where all the values smaller than $10^{-8}$ have been discarded. Notice that $m^{}_f(y^{}_f)$ in the first row actually refers to the running fermion mass $m^{}_f$ and the corresponding Yukawa coupling $y^{}_f$.}
		\label{table:coe2}
\end{table}
\end{landscape}

\begin{landscape}

\begin{table}[t]
	\scriptsize
	\centering
	\begin{tabular*}{\linewidth}{c|@{\extracolsep{\fill}}cccccccccccccccc}
		\hline
		\hline
$\left(\frac{\delta\mathcal{Y}^{}_{ik}}{\delta\mathcal{Y}^{}_{i}} \right)^2$ at $M^{}_{Z}$ & $g^{}_{s}$ & $g^{}_{}$ & $g^{\prime}_{}$ & $m^{}_{t}(y^{}_{t})$ & $m^{}_{b}(y^{}_{b})$ & $m^{}_{c}(y^{}_{c})$ & $m^{}_{s}(y^{}_{s})$ & $m^{}_{d}(y^{}_{d})$ & $m^{}_{u}(y^{}_{u})$ & $m^{}_{\tau}(y^{}_{\tau})$ & $m^{}_{\mu}(y^{}_{\mu})$ & $m^{}_{e}(y^{}_{e})$ & $v$ & $\lambda$ & $-m^2$  \\
\hline
$\alpha^{}_{s}(M^{}_{Z})$& 0.99& 0.00027& 0.00014& 0.0023& 0.38& 0.36& 0.014& 0.022& 0.0035& 0.0034& 0.0018& 0.00012& 0.0039& 0.0057& 0.0013\\
$\alpha^{-1}_{0}$& 0& 0& 0& 0& 0& 0& 0& 0& 0& 0& 0& 0& 0& 0& 0\\
$m^{}_{b}(m^{}_{b})$& 0& 0& 0& 0& 0.61& 3e-05& 0& 0& 0& 0& 0& 0& 0& 0& 0\\
$m^{}_{c}(m^{}_{c})$& 0& 0& 0& 0& 0& 0.63& 0& 0& 0& 0& 0& 0& 0& 0& 0\\
$m^{}_{s}(2~{\rm GeV})$& 0& 0& 0& 0& 0& 0& 0.99& 0& 0& 0& 0& 0& 0& 0& 0\\
$m^{}_{d}(2~{\rm GeV})$& 0& 0& 0& 0& 0& 0& 0& 0.98& 0& 0& 0& 0& 0& 0& 0\\
$m^{}_{u}(2~{\rm GeV})$& 0& 0& 0& 0& 0& 0& 0& 0& 1& 0& 0& 0& 0& 0& 0\\
$M^{}_{\tau}$& 0& 0& 0& 0& 0& 0& 0& 0& 0& 0.18& 0& 0& 0& 0& 0\\
$M^{}_{\mu}$& 0& 0& 0& 0& 0& 0& 0& 0& 0& 0& 0& 0& 0& 0& 0\\
$M^{}_{e}$& 0& 0& 0& 0& 0& 0& 0& 0& 0& 0& 0& 0& 0& 0& 0\\
$\Delta \alpha^{(5)}_{\rm had} (M^{}_{Z})$& 0& 0.0014& 0.076& 0& 0& 0& 0& 0& 0& 5.3e-05& 7e-05& 1.2e-05& 0& 0& 0\\
$G^{}_{\rm F}$& 0& 0& 0& 0& 0& 0& 0& 0& 0& 0& 0& 0& 0& 0& 0\\
$M^{}_{Z}$& 0& 0.0057& 0.0025& 0& 0& 0& 0& 0& 0& 0& 0& 0& 0& 0& 0\\
$M^{}_{h}$& 0& 0& 0& 0& 0& 0& 0& 0& 0& 0.00024& 0.00012& 0& 0.00029& 0.37& 0.63\\
$M^{}_{t}$& 4.9e-05& 0.009& 0.0088& 0.92& 7.8e-05& 2.3e-05& 0& 0& 0& 0.65& 0.33& 0.019& 0.79& 0.55& 0.31\\
\hline
EW matching& 0.015& 0& 0& 3.5e-05& 0.011& 0.00094& 0.0001& 0.00015& 2.4e-05& 0.00069& 0.00028& 1.1e-05& 5.8e-05& 8.6e-05& 2e-05\\
$b$ and $c$ decoupling, QCD& 0& 0& 0& 0& 0& 0& 0& 0& 0& 0& 0& 0& 0& 0& 0\\
$b$ and $c$ decoupling, others& 0& 0& 0& 0& 0& 0& 0& 0& 0& 0& 2.2e-05& 0& 0& 0& 0\\
$\tau$ matching& 0& 0& 0& 0& 0& 0& 0& 0& 0& 0& 5.9e-05& 0& 0& 0& 0\\
RGEs of EFT, pure QCD& 0& 0& 0& 0& 0.00019& 0.0082& 9e-05& 0.00014& 2.2e-05& 0& 0& 0& 0& 0& 0\\
RGEs of EFT, others& 0& 0& 0& 0& 0& 2.4e-05& 0& 0& 0& 5.9e-05& 0& 1.8e-05& 0& 0& 0\\
RGEs of SM, gauge couplings& 0& 0& 0& 0& 0& 0& 0& 0& 0& 0& 0& 0& 0& 0& 0\\
RGEs of SM, others& 0& 0.0021& 0.00098& 0.0017& 2.2e-05& 0& 0& 0& 0& 0.079& 0.04& 0.0023& 0.096& 0.00041& 0.00081\\
$M^{}_{\tau} \leftrightarrow m^{}_{\tau}$& 0& 0& 0& 0& 0& 0& 0& 0& 0& 5.1e-05& 0& 0& 0& 0& 0\\
$M^{}_{\mu},M^{}_{e} \leftrightarrow m^{}_{\mu}, m^{}_{e}$& 0& 0& 0& 0& 0& 0& 0& 0& 0& 0& 0.58& 0.98& 0& 0& 0\\
$M^{}_{h}$ correction& 0& 0& 0& 0& 0& 0& 0& 0& 0& 1.2e-05& 0& 0& 1.5e-05& 0.019& 0.032\\
$M^{}_{t}$ QCD correction& 0& 0.00051& 0.00022& 0.058& 0& 0& 0& 0& 0& 0.041& 0.021& 0.0012& 0.049& 0.034& 0.019\\
$M^{}_{t}$ other correction& 0& 0.00019& 8.4e-05& 0.022& 0& 0& 0& 0& 0& 0.015& 0.0077& 0.00046& 0.018& 0.013& 0.0072\\
$M^{}_{Z}$ correction& 0& 0.96& 0.43& 0& 0& 0& 0& 0& 0& 7.3e-05& 3.7e-05& 0& 0& 0& 0\\
$G^{}_{\rm F} \leftrightarrow v$& 0& 0.01& 0.0047& 4.6e-05& 1.1e-05& 0& 0& 0& 0& 0.035& 0.018& 0.001& 0.043& 0.00042& 0\\
$\alpha^{}_{0} \leftrightarrow \alpha(\mu)$& 0& 0.0086& 0.48& 0& 0& 0& 0& 0& 0& 0.00033& 0.00044& 7.3e-05& 0& 0& 0\\
		\hline
		\hline
	\end{tabular*}
	\caption{The error fractions of the output running parameters at $\mu = M^{}_{Z}$ contributed from each error source in the full SM, where the contributions smaller than $10^{-5}$ have been discarded. Notice that $m^{}_f(y^{}_f)$ in the first row actually refers to the running fermion mass $m^{}_f$ and the corresponding Yukawa coupling $y^{}_f$.}
		\label{table:errorSourcesAbove}
\end{table}
\end{landscape}

\end{document}